\documentclass{moriond}

\usepackage{amsmath}

\def\Journal#1#2#3#4{{#1} {\bf #2}, #3 (#4)}

\def\PLB{{\em Phys. Lett.}  B}
\def\PRL{\em Phys. Rev. Lett.}
\def\PRD{{\em Phys. Rev.} D}

\def\JHEP{{\em J. High Energy
Phys.}}
\def\PTEP{{\em Prog. Theor.
Exp. Phys.}}

\def\be{\begin{equation}}
\def\ee{\end{equation}}
\def\bea{\begin{eqnarray}}
\def\eea{\end{eqnarray}}

\newcommand{\Photo}{}

\begin{document}

\vspace*{4cm}
\title{Measurements of $\boldsymbol{B\rightarrow K\pi}$ and $\boldsymbol{B\rightarrow \pi\pi}$ Branching Fractions and $\boldsymbol{\mathcal{A}_{CP}}$ Asymmetries at Belle II}

\author{Shu-Ping Lin on behalf of the Belle II Collaboration}

\address{University of Padova and INFN, Via Marzolo 8, 35131, Padova, PD, Italy}

\maketitle
\abstracts{
Analyses of $B$ meson decays to charmless hadronic final states are an important part of the Belle II program. They are sensitive to effects from non-standard model physics and provide experimentally precise constraints on the weak interactions of quarks.
We present recent Belle II results on branching fractions and direct $CP$-violating asymmetries of the decays $B^0\rightarrow K^+\pi^-$, $B^+\rightarrow K^+\pi^0$, $B^+\rightarrow K^0\pi^+$, and $B^0\rightarrow K^0\pi^0$, and use these to test the standard model through an isospin-based sum rule. In addition, we measure the branching fraction and direct $CP$ asymmetry of the decay $B^+\rightarrow \pi^+\pi^0$ and the branching fraction of the decay $B^0\rightarrow \pi^+\pi^-$, which contribute towards the determination of the Cabibbo–Kobayashi–Maskawa angle $\phi_2$.
The data are collected with the Belle II detector from the SuperKEKB asymmetric-energy $e^+e^-$ collider, consisting of $387 \times 10^6$ $\Upsilon(4S)\rightarrow B\bar{B}$ events.
We obtain $-0.03 \pm 0.13 \pm 0.04$ for the sum rule, in agreement with the standard model expectation of zero and with a precision comparable to the best existing determinations.
}

\section{Motivation}

\subsection{$B\rightarrow K\pi$ Decays: Isospin Sum Rule}

For $B\rightarrow K\pi$ decays, one may construct a sum rule exploiting isospin symmetry using linear combinations of the branching fractions and $CP$ asymmetries~\cite{Gronau_2005}. This reduces the impact of theoretical uncertainties. 
The sum rule parameter
\begin{equation}
I_{K\pi} =  \mathcal{A}_{K^+\pi^-} + \mathcal{A}_{K^0\pi^+} \cdot \frac{\mathcal{B}_{K^0\pi^+}}{\mathcal{B}_{K^+\pi^-}}\frac{\tau_{B^0}}{\tau_{B^+}} 
  - 2\mathcal{A}_{K^+\pi^0} \cdot \frac{\mathcal{B}_{K^+\pi^0}}{\mathcal{B}_{K^+\pi^-}}\frac{\tau_{B^0}}{\tau_{B^+}} 
 -2\mathcal{A}_{K^0\pi^0}
\cdot\frac{\mathcal{B}_{K^0\pi^0}}{\mathcal{B}_{K^+\pi^-}}
\label{eq:sum_rule}
\end{equation}
is predicted to be zero in the Standard Model (SM) within $1\%$~\cite{Gronau_1999,Bell_2015,Bell_2020}, therefore providing a null test for the SM.
$\mathcal{A}_{K\pi}$ and $\mathcal{B}_{K\pi}$ (neutral or charged $K$ and $\pi$) are the direct $CP$ asymmetry and the $CP$-averaged branching fraction of a $B\rightarrow K\pi$ decay, and $\tau_{B^0}$ and $\tau_{B^+}$ are the neutral and charged $B$ meson lifetimes, respectively. Charge conjugation is implied unless otherwise indicated.
The time-integrated $CP$ asymmetry parameter is defined as
\begin{equation}
\mathcal{A}_{X} = \frac{\Gamma(\bar{B}\rightarrow\bar{X}) - \Gamma(B\rightarrow X)}{\Gamma(\bar{B}\rightarrow\bar{X}) + \Gamma(B\rightarrow X)},
\end{equation}
where $\Gamma$ is the partial decay width to the final state $X$.
$\mathcal{A}_{X}$ corresponds to the direct $CP$ asymmetry of charged $B$ mesons and, in the absence of $CP$ violation in flavor oscillations, also neutral $B$ mesons. This is a good approximation in the case of $B^0$-$\bar{B}^0$ mixing.

The current world average of the sum rule parameter is $I_{K\pi} = -0.13 \pm 0.11$~\cite{Amhis_2023}, and its precision is limited by the statistical uncertainty of $\mathcal{A}_{K^0\pi^0}$~\cite{Workman_2022}.

\subsection{$B\rightarrow \pi\pi$ Decays: Towards the Angle $\phi_2$}
The precision of the angle $\phi_2$ (or $\alpha$) $ = \arg( - V_{td}V^*_{tb}/V_{ud}V_{ub}^* )$,
where $V_{ij}$ are the elements of the Cabibbo–Kobayashi–Maskawa (CKM) quark-mixing matrix~\cite{Cabibbo_1963,Kobayashi_1973}, is a limiting factor of the global testing power of the CKM model.
$\phi_2$ is accessible through the combined information of $B^0\rightarrow \pi^+\pi^-$, $B^+\rightarrow\pi^+\pi^0$, and $B^0\rightarrow\pi^0\pi^0$ decays~\cite{Gronau_1990,Charles_2017}.
Precise measurements of the branching fraction and $CP$ asymmetries of the decays $B^0\rightarrow \pi^+\pi^-$ and $B^+\rightarrow\pi^+\pi^0$ provide constraints through isospin symmetry relations in the determination of $\phi_2$.

\section{Analysis Strategy}
The analyses of all the decays follow a similar strategy. A common selection is applied to all relevant final state particles. Multivariate algorithms are developed for each mode to suppress the continuum background, which originates from $e^+e^-\rightarrow q\bar{q}$ processes, where $q$ indicates a $u$, $d$, $s$, or $c$ quark. 
Event shape variables, results of the $B$ meson decay vertex fit, flavor-tagger outputs (except for $B^0\rightarrow K_S^0\pi^0$ decay), and the beam-constrained mass $M_{bc} = \sqrt{ (E_{beam}^*/c^2)^2 - (p_{B}^*/c)^2 }$ are used to train a boosted decision tree (BDT). $E_{beam}^*$ and $p_B^*$ are the beam energy and $B$ meson momentum, respectively, in the center-of-mass frame. A loose requirement is applied to the BDT output, removing $90\%-99\%$ of the continuum background.
In order to improve the $M_{bc}$ resolution for decays containing a neutral $\pi^0$, the measured $\pi^0$ momentum in the calculation of $M_{bc}$ is replaced by
\begin{equation}
    \boldsymbol{p}_{\pi^0}^{*'} = \sqrt{ (E_{beam}^* - E_h^*)^2/c^2 - m_{\pi^0}^2c^2 } \times \frac{\boldsymbol{p}_{\pi^0}^*}{|\boldsymbol{p}_{\pi^0}^*|},
\end{equation}
where $E_h^*$ is energy of the other hadron in the $B$ candidate decay, $\boldsymbol{p}_{\pi^0}^*$ is the measured $\pi^0$ momentum, and $m_{\pi^0}$ is the known $\pi^0$ mass.
This substitution improves the resolution of $M_{bc}$ and reduces its correlation to $\Delta E$.

The signal yields are measured with unbinned maximum likelihood fits to the energy difference $\Delta E = E_B^* - E_{beam}^*$ and the transformed BDT output $C'$.
$E_B^*$ is the energy of the $B$ meson in the center-of-mass frame, and the BDT output is transformed using probability integral transformation~\cite{muTrans}.
The two-dimensional probability density functions (PDF) of $\Delta E$ and $C'$ is assumed to be a product of two independent one-dimensional PDFs, as the correlation between the two variables is small.
The $CP$ asymmetries are determined using the charge of the final state particles for flavor-specific channels.
For the $B^0\rightarrow K_S^0\pi^0$ channel the $B$ mesons decay to the same $CP$ eigenstate, and a category-based flavor tagger is used to determine the flavor of the $B^0$, calibrated using $B\rightarrow D^{(*)} h^+$ decays~\cite{flv_tg}.

The analysis procedures are first studied with simulated events before being applied to the data sample. 
Several abundant control channels with similar features to the signal channels are used to correct for discrepancies between data and simulation, and to estimate systematic uncertainties.

\section{Result}
\begin{figure}[htb]
\includegraphics[width=0.49\linewidth]{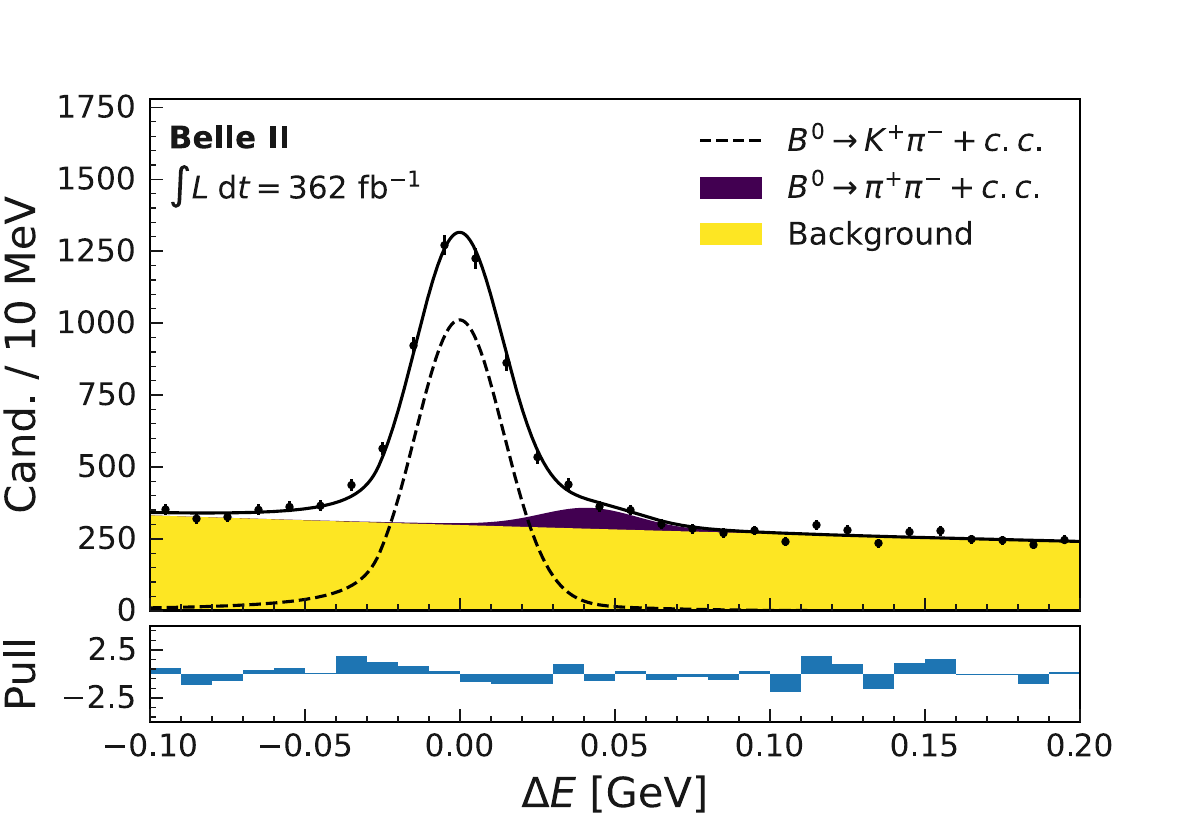}
\includegraphics[width=0.49\linewidth]{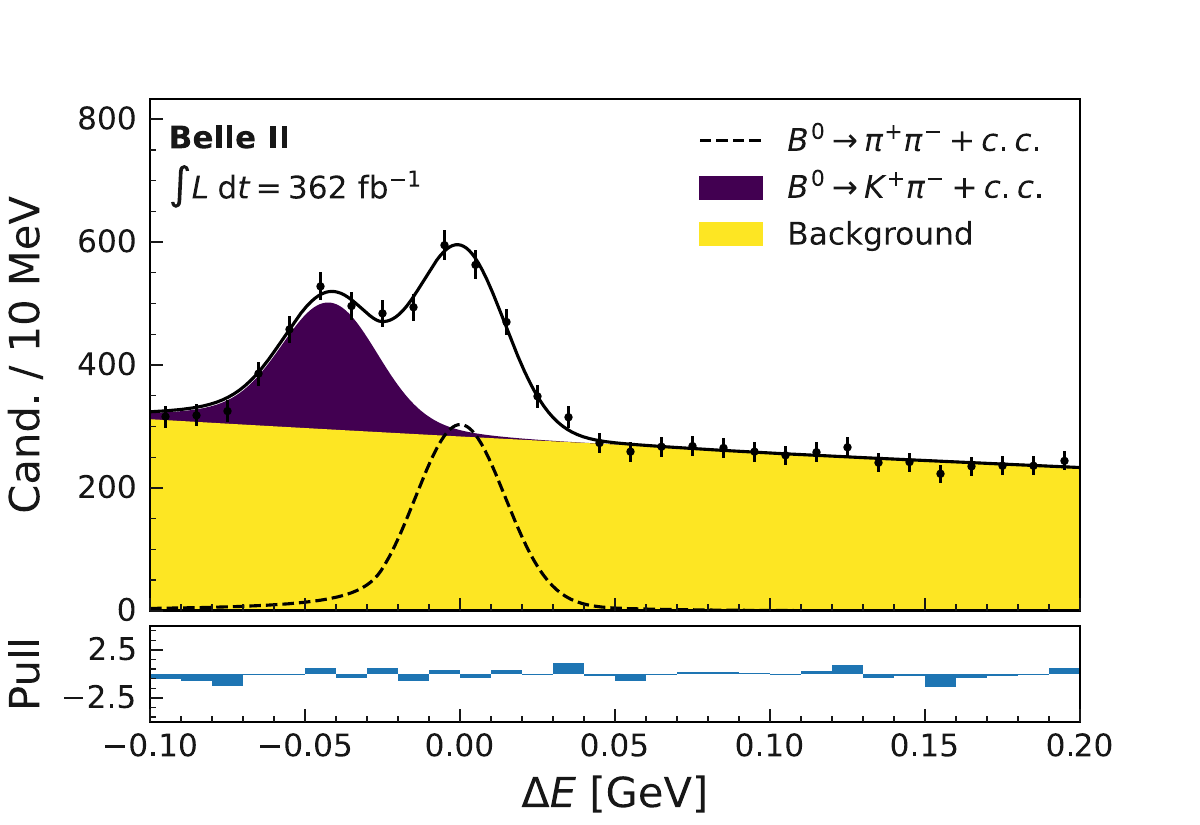}
\includegraphics[width=0.49\linewidth]{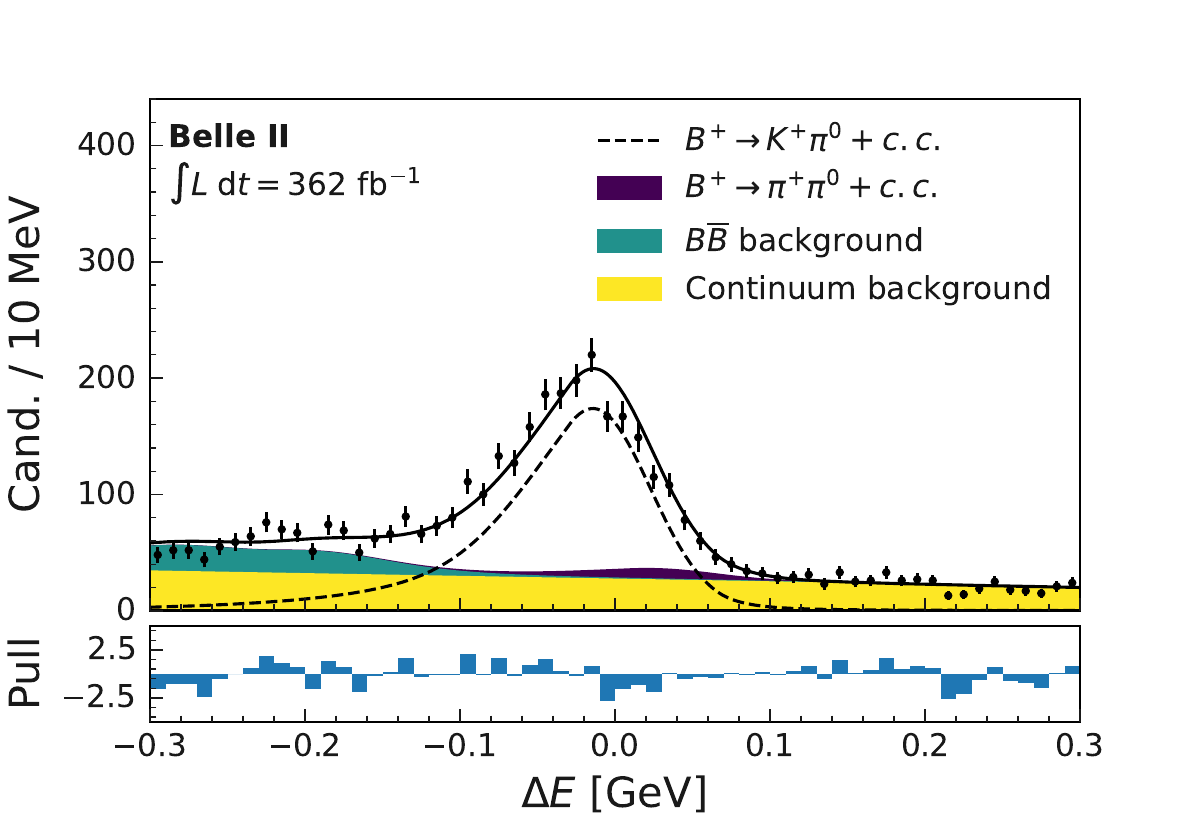}
\includegraphics[width=0.49\linewidth]{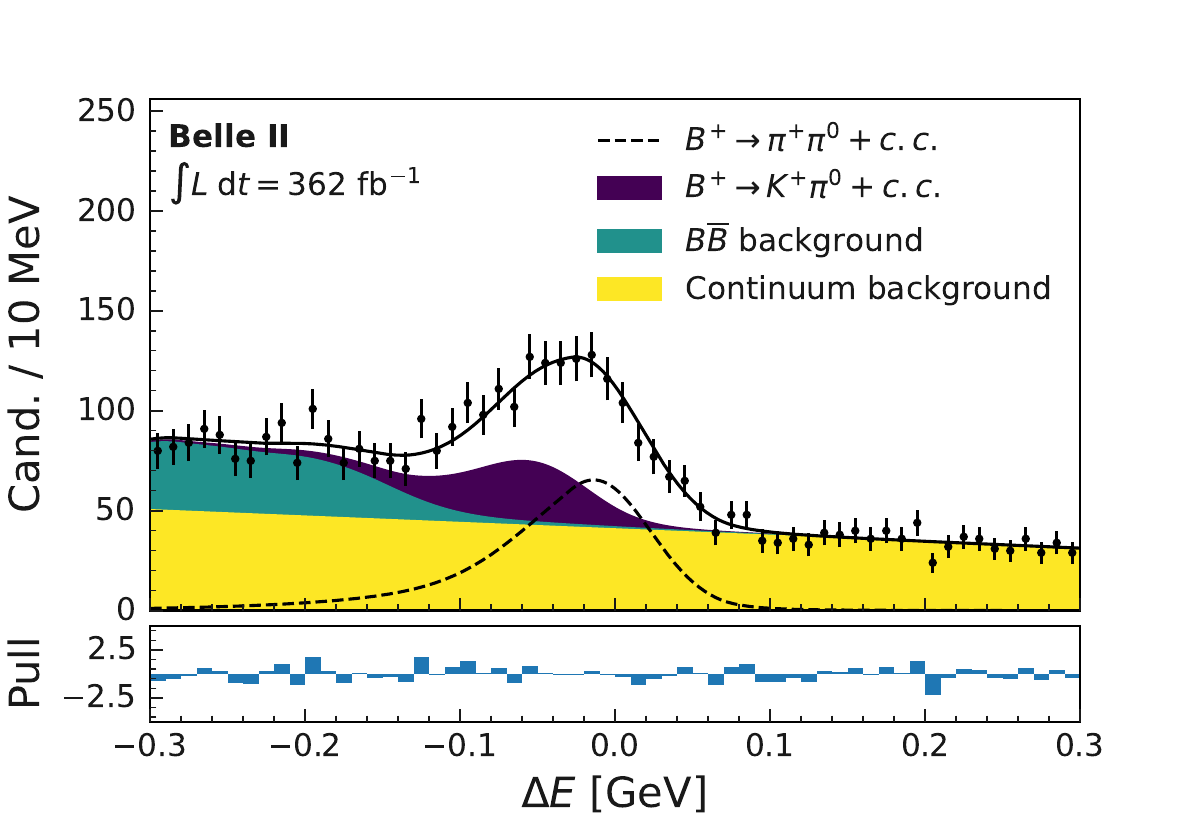}
\includegraphics[width=0.49\linewidth]{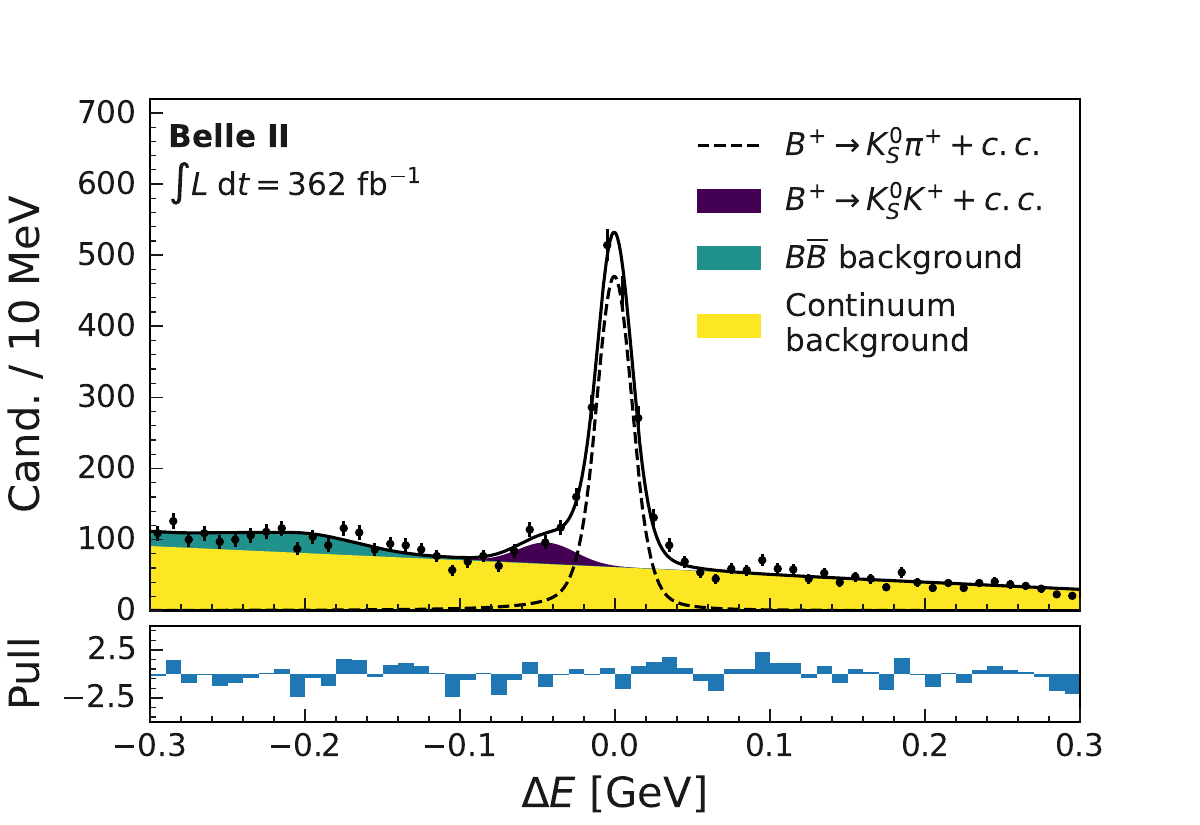}
\includegraphics[width=0.49\linewidth]{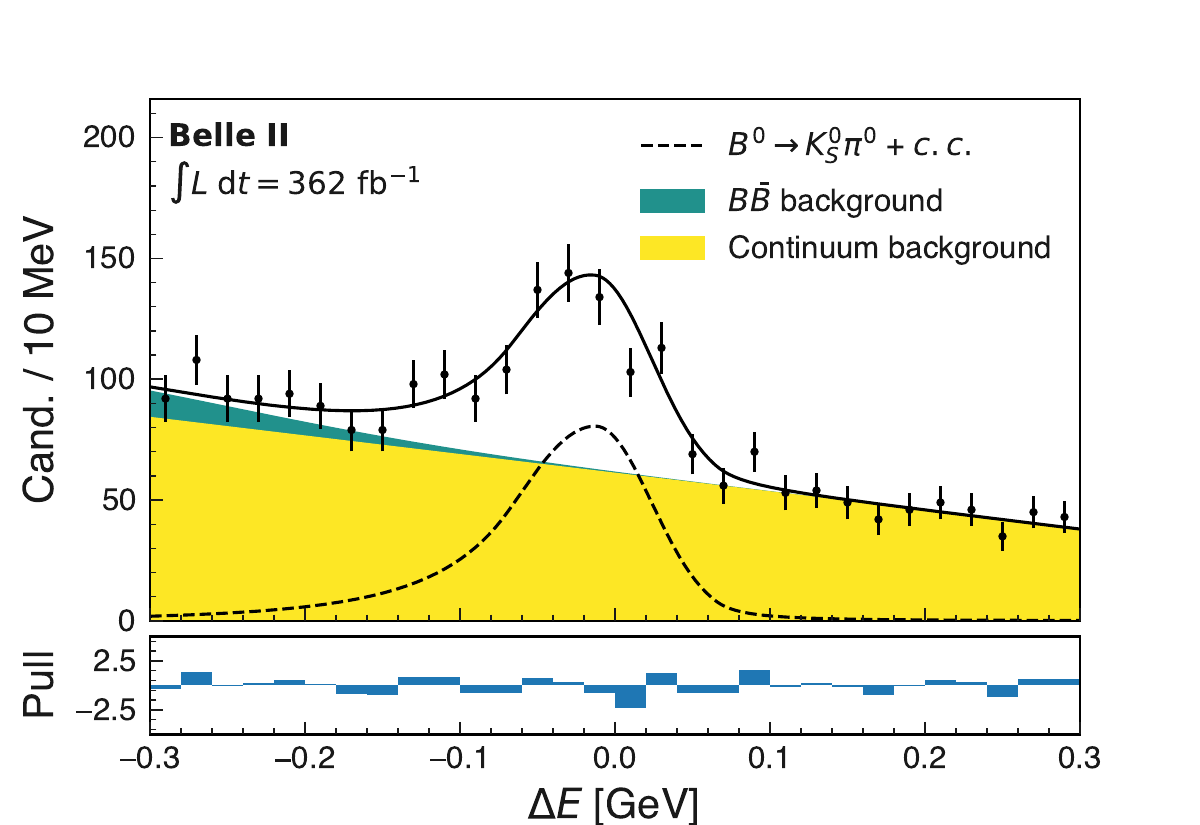}
\caption[]{$\Delta E$ fit projections and data distributions for $B^0\rightarrow K^+\pi^-$ (upper left), $B^0\rightarrow \pi^+\pi^-$ (upper right), $B^+\rightarrow K^+\pi^0$ (middle left), $B^+\rightarrow \pi^+\pi^0$ (middle right), $B^+\rightarrow K_S^0\pi^+$ (lower left) and $B^0\rightarrow K_S^0\pi^0$ (lower right) decays.}

\label{fig:deltaE}
\end{figure}

The fit results are listed in Table~\ref{tab:RST}, and the fit projections to $\Delta E$ are shown in Figure~\ref{fig:deltaE}.
All results~\cite{Btohh} agree with the world averages, and our precisions are comparable to the best existing determinations from Belle~\cite{Duh_2013,Dalseno_2013,Fujikawa_2010} and BABAR~\cite{Lees_2013,Aubert_2007_75,Aubert_2007_76,Aubert_2009} despite using a smaller data sample.

The branching fraction and direct $CP$ asymmetry of $B^0\rightarrow K^0\pi^0$ decays have also been measured in an analysis of the decay time evolution~\cite{kspiz_td} based on the same data sample, but featuring a different event selection. 
The result presented in Table~\ref{tab:RST} is the combination of both analyses, where the statistical and systematic correlations are taken into account. The result for the $CP$ asymmetry of $B^0\rightarrow K^0\pi^0$ and the branching fraction of $B^0\rightarrow\pi^+\pi^-$ are the most precise measurement made by a single experiment to date.

The sum rule of Equation~\ref{eq:sum_rule} is calculated using our measurements of the branching fractions and $CP$ asymmetries,
\begin{equation}
    I_{K\pi} = -0.03 \pm 0.13\pm0.04,
\end{equation}
where the measured ratio $\tau_{B^0}/\tau_{B^+}=0.9273\pm0.0033$~\cite{Workman_2022} is used, and the correlations between uncertainties are taken into account. Common systematic uncertainties are cancelled in the ratios of the branching fractions in Equation~\ref{eq:sum_rule}.
The sum rule result is consistent with the theoretical expectation. It is statistically limited and has a similar precision to that from the average of measurements from Belle, BABAR and LHCb collaborations~\cite{Amhis_2023}.

\begin{table}[htb]
    \centering
    \caption{Results of the branching fraction and $CP$ asymmetries using $362\text{ fb}^{-1}$ Belle II data. The first uncertainties are statistical and the second are systematic.}
    \label{tab:RST}
    \vspace{0.2cm}
    \begin{tabular}{|c|c|c|}
    \hline
    & $\mathcal{B}\ [10^{-6}]$ & $\mathcal{A}_{{\it CP}}$ \\
    \hline
    $B^0\rightarrow K^+\pi^-$ & $20.67\pm0.37\pm0.62$ & $-0.072\pm0.019\pm0.007$ \\
    $B^0\rightarrow \pi^+\pi^-$ & $5.83\pm0.22\pm0.17$ & -- \\
    
    $B^+\rightarrow K^+\pi^0$ & $13.93\pm0.38\pm0.71$ & $0.013\pm0.027\pm0.005$ \\
    $B^+\rightarrow \pi^+\pi^0$ & $5.10\pm0.29\pm0.27$ & $-0.081\pm0.054\pm0.008$ \\
    
    $B^+\rightarrow K^0 \pi^+$ & $24.37\pm0.71\pm0.86$ & $0.046\pm0.029\pm0.007$\\
    $B^0\rightarrow K^0 \pi^0$ & $10.73\pm0.63\pm0.62$ & $-0.01\pm0.12\pm0.04$ \\
    \hline
    \end{tabular}
\end{table}



\end{document}